\documentclass[twocolumn,showpacs,preprintnumbers,amsmath,amssymb]{revtex4}

\usepackage{graphicx}
\usepackage{dcolumn}
\usepackage{bm}

\begin{document}

\preprint{APS/123-QED}

\title{Type I background fields in terms of type IIB ones}

\author{B. Nikoli\'c}
 \email{bnikolic@phy.bg.ac.yu}
\author{B. Sazdovi\'c}%
 \email{sazdovic@phy.bg.ac.yu}
\affiliation{%
Institute of Physics, 11001 Belgrade, P.O.Box 57, Serbia
}%

\date{\today}

\begin{abstract}
We choose such boundary conditions for open IIB superstring theory
which preserve $N=1$ SUSY. The explicite solution of the boundary
conditions yields effective theory which is symmetric under
world-sheet parity transformation $\Omega:\sigma\to-\sigma$. We
recognize effective theory as closed type I superstring theory.
Its background fields,beside known $\Omega$ even fields of the
initial IIB theory, contain improvements quadratic in $\Omega$ odd
ones.
\end{abstract}

\maketitle

\section{\label{sec:level1}Introduction}

There are five consistent supersymmetric string theories: type I,
type IIA and IIB and two heterotic string theories. They are
related by web of dualities and in fact provide different
descriptions of the same theory \cite{PBBS}. In this article we
are going to improve known relation between type I and type IIB
theories.

It is known that type I superstring theory can be obtained from
type IIB superstring theory as a projection on the states that are
even under world-sheet parity transformation
$\Omega:\sigma\to-\sigma$. After this orientifold projection the
following background fields survive: graviton $G_{\mu\nu}$ and
dilaton $\Phi$ from NS-NS sector, the sum of two same chirality
gravitions
$\psi^\alpha_{+\mu}=\psi^\alpha_\mu+\bar\psi^\alpha_\mu$ and
dilatinos $\lambda^\alpha_{+}=\lambda^\alpha+\bar\lambda^\alpha$
from NS-R sector and two rank antisymmetric tenzor $C_{\mu\nu}$
from R-R sector. The states that are odd under $\Omega$
transformations: antisymmetric tensor $B_{\mu\nu}$ from NS-NS
sector, difference of two gravitinos
$\psi^\alpha_{-\mu}=\psi^\alpha_\mu-\bar\psi^\alpha_{\mu}$ and dilatinos
$\lambda^\alpha_{-}=\lambda^\alpha -\bar\lambda^\alpha$
from NS-R sector, and scalar $C_0$ and four rank
antisymmetric tensor $C_{\mu\nu\rho\sigma}$ with self dual field
strength from R-R sector, are eliminated by above projection.

We are going to investigate Green-Schwarz formulation of type IIB
superstring adopted by Refs.\cite{berko}. In particular, we use
the form of Ref.\cite{susyNC} which corresponds to constant
graviton $G_{\mu\nu}$, antisymmetric field $B_{\mu\nu}$, two
gravitinos $\psi^\alpha_\mu$ and $\bar\psi^\alpha_\mu$, and R-R
field strength $F^{\alpha\beta}$. In this approach dilaton $\Phi$
and two dilatinos $\lambda^\alpha$ and $\bar\lambda^\alpha$ are
set to zero.

Let us express connection between two descriptions of R-R sector
\cite{PBBS}. It is known that bispinor $F^{\alpha\beta}=i S^\alpha
(\Gamma^0 \tilde S)^\beta$, made from same chirality spinors
$S^\alpha$ and $\tilde S^{\alpha}$, can be expanded into complete
set of antisymmetric gamma matrices
\begin{equation}
F^{\alpha\beta}=\sum_{k=0}^D
\frac{i^k}{k!}F_{(k)}\Gamma_{(k)}^{\alpha\beta}\, ,  \quad \left[
\Gamma_{(k)}^{\alpha\beta}=(\Gamma^{[\mu_1\dots
\mu_k]})^{\alpha\beta}\right]
\end{equation}
where in short-hand notation $F_{(k)}$ is $k$-rank antisymmetric
tensor. As a consequence of chirality conditions on bispinor
$F^{\alpha\beta}$ and duality relations, the independent tensors
are $F_{(1)}$, $F_{(3)}$ and self-dual part of $F_{(5)}$. Using
physical state condition these tensors can be solved in terms of
potentials $F_{(k)}=dC_{(k-1)}$, so that IIB theory contains just
the potentials $C_{(0)}$, $C_{(2)}$ and $C_{(4)}$. Note that
symmetric part of $F^{\alpha\beta}$,
$F_s^{\alpha\beta}=\frac{1}{2}(F^{\alpha\beta}+F^{\beta\alpha})$,
corresponds to the potentials $C_0$ and $C_{\mu\nu\rho\sigma}$,
and antisymmetric part,
$F_a^{\alpha\beta}=\frac{1}{2}(F^{\alpha\beta}-F^{\beta\alpha})$,
corresponds to the potential $C_{\mu\nu}$.

We start with type IIB open superstring theory
described by bosonic coordinates $x^\mu$,  and same chirality
fermionic ones $\theta^\alpha$ and $\bar\theta^\alpha$.
According to Refs.\cite{BNBS}-\cite{radepjc} we apply the
canonical method, treating boundary conditions as the canonical constraints. We choose
Neumann boundary conditions for bosonic coordinates $x^\mu$ and
$(\theta^\alpha-\bar\theta^\alpha)|_0^\pi=0$ for fermionic
coordinates in order to preserve $N=1$ SUSY from initial $N=2$
SUSY in type IIB theory. We are able to solve all boundary
conditions and obtain effective theory. It occurs that this
effective theory, even under world-sheet parity transformation
$\Omega$, is just type I closed superstring theory defined on
orientifold projection. The main result of the article contains generalized expressions
of type I superstring background fields in terms of type IIB ones
\begin{eqnarray}\label{eq:sub2b1}
G_{\mu\nu}&\to&
G_{\mu\nu}^{eff}=G_{\mu\nu}-4B_{\mu\rho}G^{\rho\sigma}
B_{\sigma\nu}\, ,\,\, B_{\mu\nu}\to 0\, ,\nonumber \\
\frac{1}{2}\psi^\alpha_{+\mu}&\to&
(\Psi_{eff})^\alpha_\mu=\frac{1}{2}\psi^\alpha_{+\mu}+B_{\mu\rho}G^{\rho\nu}\psi^\alpha_{-\nu}\,
,\,\,\psi^\alpha_{-\mu}\to0\, , \nonumber \\
F_{\mu\nu\rho}&\to&
F_{\mu\nu\rho}^{eff}=F_{\mu\nu\rho}-\frac{1}{D}G^{\varepsilon\eta}
\psi_{-\varepsilon}\Gamma_{[\mu\nu\rho]} \psi_{-\eta}\, ,\nonumber\\F_\mu&\to& 0\,
,\quad F_{\mu\nu\rho\sigma\varepsilon}\to 0\, .
\end{eqnarray}
Therefore, fields $B_{\mu\nu}$ and $\psi^\alpha_{-\mu}$, which describe
states odd under world-sheet parity transformations, are not complitely
eliminated, but they survive as quadratic terms in type I
superstring background.

We will refer to the fields $G_{\mu\nu}$, $B_{\mu\nu}$, $\psi^\alpha_\mu$,
$\bar\psi^\alpha_\mu$ and $F^{\alpha\beta}$ ( or
$F_{(1)}$, $F_{(3)}$ and $F_{(5)}$) as
\textit{type IIB} background fields
and to fields $G_{\mu\nu}^{eff}$,
$(\Psi_{eff})^\alpha_\mu$ and $F^{\alpha\beta}_{eff}$ (or $F_{(3)}^{eff}$ )
as \textit{type I} background fields (the
background fields seen by type IIB and type I superstring
respectively). The names are taken from initial and final
(effective) descriptions of the same theory.

\section{\label{sec:level1}Canonical approach to Green-Schwarz formulation of IIB theory}

We will investigate the type IIB superstring theory using
Green-Scwarz  formulation of Ref.\cite{susyNC}. The action is
\begin{eqnarray}\label{eq:SB}
&S&=\kappa \int_\Sigma d^2\xi \left[
\frac{1}{2}\eta^{ab}G_{\mu\nu}+\varepsilon^{ab} B_{\mu\nu}\right]\partial_a
x^\mu \partial_b x^\nu\nonumber \\&-&\int_\Sigma d^2 \xi  \pi_\alpha
(\partial_\tau-\partial_\sigma)(\theta^\alpha+\psi^\alpha_\mu
x^\mu)\nonumber\\&+&\int_\Sigma d^2 \xi\left[ (\partial_\tau+\partial_\sigma)(\bar\theta^{\alpha}+\bar
\psi^{\alpha}_\mu x^\mu)\bar\pi_{\alpha}+\frac{1}{2\kappa}\pi_\alpha F^{\alpha
\beta}\bar \pi_{\beta}\right] \, \,\,\,
\end{eqnarray}
where the world sheet $\Sigma$ is parameterized by
$\xi^a=(\xi^0=\tau\, ,\xi^1=\sigma)$, and $D=10$ dimensional
space-time is parameterized by
coordinates $x^\mu$ ($\mu=0,1,2,\dots,D-1$). The
fermionic part of superspace is spanned by same chirality
fermionic coordinates $\theta^\alpha$ and
$\bar\theta^{\alpha}$, while the variables $\pi_\alpha$ and $\bar \pi_{\alpha}$ are their canonically conjugated
momenta.

\subsection{\label{sec:level2}Canonical Hamiltonian and boundary conditions as canonical constraints}

According to the definition of canonical Hamiltonian, $\mathcal H_c=\dot x^\mu \pi_\mu+\dot
\theta^\alpha \pi_\alpha+\dot
{\bar\theta}^{\alpha}\bar\pi_{\alpha}-\mathcal L$, we have
\begin{equation}\label{eq:initialham}
H_c=\int d\sigma \mathcal H_c\, ,\quad \mathcal H_c=T_--T_+\, ,\quad T_{\pm} =t_{\pm}-\tau_{\pm}\, ,
\end{equation}
where
\begin{eqnarray}
t_{\pm}&=&\mp\frac{1}{4\kappa}G^{\mu\nu}I_{\pm \mu}I_{\pm \nu}\,
,\nonumber\\I_{\pm \mu}&=&\pi_\mu+2\kappa \Pi_{\pm
\mu\nu}x'^\nu+\pi_\alpha \psi^\alpha_\mu-\bar\psi^{\alpha}_\mu
\bar\pi_{\alpha}\, ,\nonumber \\
\tau_+&=&(\theta'^\alpha+\psi^\alpha_\mu x'^\mu)
\pi_\alpha-\frac{1}{4\kappa}\pi_\alpha
F^{\alpha\beta}\bar\pi_{\beta}\, ,\nonumber\\
\tau_-&=&(\bar\theta'^\alpha+\bar\psi^\alpha_\mu
x'^\mu)\bar\pi_{\alpha}+\frac{1}{4\kappa}\pi_\alpha
F^{\alpha\beta}\bar\pi_{\beta}\, , \label{eq:struja}
\end{eqnarray}
and $\pi_\mu$ denotes momentum canonically conjugated to the coordinate $x^\mu$.
Using the standard Poisson bracket algebra we find that components $T_{\pm}$ satisfy
Virasoro algebra.

Following method of Ref.\cite{radepjc}, using canonical approach,
we will derive boundary conditions directly in terms of canonical
variables. Varying Hamiltonian $H_c$ we obtain
\begin{equation}
\delta H_c=\delta H_c^{(R)}-[\gamma_\mu^{(0)}\delta x^\mu+
\pi_\alpha\delta\theta^\alpha+\delta
\bar\theta^{\alpha}\bar\pi_{\alpha}]\big |_0^\pi\, ,
\end{equation}
where $\delta H_c^{(R)}$ is regular term without $\tau$ and
$\sigma$  derivatives of coordinate variations, $\delta x^\mu$,
$\delta \theta^\alpha$, $\delta\bar\theta^{\alpha}$, and
variations of their canonically conjugated momenta, $\delta
\pi_\mu$, $\delta \pi_\alpha$ and $\delta \bar\pi_\alpha$, while
\begin{equation}
\gamma_{\mu}^{(0)}=\Pi_{+ \mu}{}^\nu I_{- \nu}+\Pi_{- \mu}{}^\nu
I_{+ \nu}+\pi_\alpha \psi^\alpha_\mu+\bar\psi^{\alpha}_\mu
\bar\pi_{\alpha}\, .
\end{equation}
As a time translation generator canonical Hamiltonian must have
well defined derivatives in its variables. Consequently, boundary
term has to vanish and we obtain
\begin{equation}\label{eq:BC}
\left[\gamma_\mu^{(0)}\delta x^\mu+
\pi_\alpha\delta\theta^\alpha+\delta
\bar\theta^{\alpha}\bar\pi_{\alpha}\right]\Big |_0^\pi=0\, .
\end{equation}

For bosonic coordinates $x^\mu$ we choose Neumann boundary conditions implying
\begin{equation}\label{eq:gamami}
\gamma_\mu^{(0)}\big |_0^\pi=0\, .
\end{equation}
In order to preserve $N=1$ SUSY of the initial $N=2$ SUSY,
following  \cite{susyNC}, for fermionic coordinates we chose
\begin{equation}\label{eq:bcf1}
(\theta^\alpha-\bar\theta^{\alpha})\Big |_0^\pi=0\, , \qquad
(\pi_\alpha-\bar\pi_{\alpha})\big |_0^\pi=0\, ,
\end{equation}
where the first condition produces the second one.
According to Refs.\cite{BNBS,radepjc},  we will treat the expressions
(\ref{eq:gamami})-(\ref{eq:bcf1}) as canonical constraints.

\subsection{\label{sec:level2} Dirac consistency procedure}

As well as in Refs.\cite{BNBS}-\cite{kanonski},
Dirac canonical consistency procedure for constraints
(\ref{eq:gamami}) gives an infinite set of constraints which can
be rewritten in the compact $\sigma$-dependent form
\begin{eqnarray}\label{eq:uslov1}
\Gamma_\mu(\sigma)&=&\Pi_{+ \mu}{}^\nu
I_{- \nu}(\sigma)+\Pi_{- \mu}{}^\nu I_{+
\nu}(-\sigma)\nonumber\\&+&\pi_\alpha(-\sigma)\psi^\alpha_{\mu}+\bar\psi^{\alpha}_\mu
\bar\pi_{\alpha}(\sigma)\, .
\end{eqnarray}

Applying the same procedure for fermionic boundary conditions
(\ref{eq:bcf1}),  we obtain respectively
\begin{eqnarray}\label{eq:uslov2}
\Gamma^\alpha(\sigma)=
\Theta^\alpha(\sigma)-\bar\Theta^{\alpha}(\sigma)\, ,\,
\Gamma_{\alpha}^\pi(\sigma)=\pi_\alpha(-\sigma)-\bar\pi_{\alpha}(\sigma)\,
\end{eqnarray}
where
\begin{eqnarray}\label{eq:Fialfa}
\Theta^\alpha(\sigma)&=&\theta^\alpha(-\sigma)-\psi^\alpha_\mu
\tilde q^\mu(\sigma)-\frac{1}{2\kappa}F^{\alpha\beta}\int_0^\sigma d\sigma_1
P_s \bar\pi_{\beta}            \nonumber\\
&+&\frac{1}{2\kappa}G^{\mu\nu} \psi^\alpha_\mu   \int_0^\sigma
d\sigma_1 P_s (I_{+ \nu}+I_{- \nu})  \, ,
\end{eqnarray}
\begin{eqnarray}\label{eq:barFialfa}
\bar\Theta^{\alpha}(\sigma)&=&\bar\theta^{\alpha}(\sigma)+\bar\psi^{\alpha}_\mu
\tilde q^\mu(\sigma)+\frac{1}{2\kappa} F^{\beta\alpha}    \int_0^\sigma d\sigma_1
P_s \pi_{\beta} \,\,   \nonumber\\
&+&\frac{1}{2\kappa}G^{\mu\nu} \bar\psi^{\alpha}_\mu   \int_0^\sigma
d\sigma_1 P_s (I_{+ \nu}+I_{- \nu})  \, .
\end{eqnarray}
We introduced new variables, symmetric and antisymmetric
under world-sheet parity transformation $\Omega:\sigma\to -\sigma$. For bosonic variables we use standard
notation \cite{BNBS}-\cite{radepjc}
\begin{eqnarray}\label{eq:bv1}
q^\mu(\sigma)&=&P_s x^\mu(\sigma)\, ,\quad \tilde
q^\mu(\sigma)=P_a x^\mu(\sigma)\, ,\nonumber\\ p_\mu(\sigma)&=&P_s \pi_\mu(\sigma)\, ,\quad \tilde
p_\mu(\sigma)=P_a \pi_\mu(\sigma)\, ,
\end{eqnarray}
while for fermionic ones we use
the projectors on $\sigma$ symmetric and antisymmetric parts
\begin{equation}
P_s=\frac{1}{2}(1+\Omega)\, ,\quad P_a=\frac{1}{2}(1-\Omega)\, .
\end{equation}

From $\left\lbrace H_c\,
,\Gamma_A\right\rbrace=\Gamma_A'\approx0\, ,\Gamma_A=(\Gamma_\mu\,
,\Gamma^\alpha\, ,\Gamma_\alpha^\pi)\, ,$ it follows that there
are no more constraints in the theory.

Also, there exists the set of constraints at $\sigma=\pi$,
$\bar\Gamma_\mu(\sigma)$,  $\bar\Gamma^{\alpha}(\sigma)$ and
$\bar\Gamma_\pi^\alpha(\sigma)$. They differ from conditions
(\ref{eq:uslov1}) and (\ref{eq:uslov2}) only in terms depending on
$-\sigma$ and can be obtained just replacing $-\sigma$ with
$2\pi-\sigma$. Constraints at $\sigma =\pi$ can be solved by
$2\pi$ periodicity of all canonical variables as well as in
Refs.\cite{BNBS,radepjc}. So, the effective theory will be
closed string theory.

\subsection{\label{sec:level2}Classification of constraints}

Computing the algebra of the constraints ${}^\star{}\Gamma_A=(\Gamma_\mu\, ,\Gamma'^\alpha\, ,\Gamma^\pi_\alpha)$
\begin{equation}
\left\lbrace {}^\star\Gamma_A\, ,{}^\star\Gamma_B\right\rbrace=M_{AB}\delta'\, ,
\end{equation}
we obtain the supermatrix
\begin{equation}
M_{AB}=\left(
\begin{array}{ccc}
-\kappa G^{eff}_{\mu\nu} & 2(\Psi_{eff})^\gamma_\mu & 0\\
-2(\Psi_{eff})^\alpha_\nu & -\frac{1}{\kappa}F^{\alpha\gamma}_{eff} & -2\delta^\alpha{}_\delta \\
0 & -2\delta_\beta{}^\gamma & 0
\end{array}\right)\, ,
\end{equation}
with the effective background fields
\begin{eqnarray}\label{eq:effbackground}
G^{eff}_{\mu\nu}&=&G_{\mu\nu}-4B_{\mu\rho}G^{\rho\lambda}B_{\lambda\nu}\, ,\nonumber\\
(\Psi_{eff})^\alpha_{\mu}&=&\frac{1}{2}\psi^\alpha_{+\mu}+B_{\mu\rho}G^{\rho\nu} \psi^\alpha_{-\nu}\,
,\nonumber \\
F_{eff}^{\alpha\beta}&=&F_{a}^{\alpha\beta}-\psi^\alpha_{-\mu}G^{\mu\nu}\psi^\beta_{-\nu}\, ,
\end{eqnarray}
where the fields $\psi^\alpha_{\pm\mu}$ are defined as
$\psi^\alpha_{\pm \mu}=\psi^\alpha_\mu\pm\bar\psi^{\alpha}_\mu$.
The superdeterminant $s\det M_{AB}$ is proportional to $\det
G^{eff}_{\mu\nu}$,  which is assumed to be different from zero.
Consequently, all constraints are of the second class and we can solve them explicitely.

\section{\label{sec:level1} Type I superstring as effective theory}

The solution of the constraint equations
$\Gamma_\mu=0$, $\Gamma^\alpha=0$ and $\Gamma_\alpha^\pi=0$ has
the form
\begin{equation}\label{eq:resenjex}
x^{\mu}(\sigma)=q^\mu-2\Theta^{\mu\nu}\int_0^\sigma d\sigma_1
p_\nu+ \frac{\Theta^{\mu\alpha}}{2}\int_0^\sigma d\sigma_1 (p_{\alpha}+{\bar p}_\alpha)\,  ,\nonumber
\end{equation}
\begin{eqnarray}\label{eq:resenje1}
\theta^\alpha(\sigma)&=&\eta^\alpha-\Theta^{\mu\alpha}\int_0^\sigma
d\sigma_1 p_\mu -\frac{\Theta^{\alpha\beta}}{4}\int_0^\sigma d\sigma_1
(p_\beta +{\bar p}_\beta)   \, ,\nonumber
\end{eqnarray}
\begin{eqnarray}\label{eq:resenje2}
\bar\theta^\alpha(\sigma)&=&\bar\eta^\alpha-\Theta^{\mu\alpha}\int_0^\sigma
d\sigma_1 p_\mu - \frac{\Theta^{\alpha\beta}}{4}\int_0^\sigma d\sigma_1
(p_\beta +{\bar p}_\beta)\, ,\nonumber
\end{eqnarray}
\begin{eqnarray}\label{eq:resenje2}
\pi_\mu=p_\mu\, , \qquad  \pi_\alpha&=& \frac{p_\alpha}{2}  \, , \qquad \bar\pi_{\alpha}= \frac{{\bar p}_\alpha}{2}\, ,
\end{eqnarray}
where
\begin{eqnarray}\label{eq:resenjepi}
\eta^\alpha&\equiv&\frac{1}{2}(\theta^\alpha+ \Omega
\bar\theta^\alpha)\, ,\quad \bar\eta^\alpha\equiv \frac{1}{2}(\Omega
\theta^\alpha+ \bar\theta^\alpha) \, ,\nonumber\\
p_\alpha&\equiv&\pi_\alpha+\Omega \bar\pi_{\alpha} \, ,\quad
\bar p_\alpha\equiv \Omega \pi_\alpha+ \bar\pi_{\alpha}  \, ,
\end{eqnarray}
and
\begin{eqnarray}
\Theta^{\mu\nu}&=&-\frac{1}{\kappa}(G_{eff}^{-1}BG^{-1})^{\mu\nu}\,
,\nonumber\\
\Theta^{\mu\alpha}&=&2\Theta^{\mu\nu}(\Psi_{eff})^\alpha_{\nu}-\frac{1}{2\kappa}G^{\mu\nu}\psi^\alpha_{-\nu}\,
,  \nonumber\\
\Theta^{\alpha\beta}&=&\frac{1}{2\kappa}F_s^{\alpha\beta}+4(\Psi_{eff})^\alpha_{\mu}
\Theta^{\mu\nu}(\Psi_{eff})^\beta_{\nu}\nonumber\\
&-&\frac{1}{\kappa}\psi^\alpha_{-\mu}(G^{-1}BG^{-1})^{\mu\nu}\psi^\beta_{-\nu}\nonumber \\ &+&\frac{G^{\mu\nu}}{\kappa}\left[ \psi^\alpha_{-\mu}(\Psi_{eff})^\beta_{\nu}+\psi^\beta_{-\mu}(\Psi_{eff})^\alpha_{\nu}
\right] \,
.
\end{eqnarray}

Substituting the solutions of the constraints (\ref{eq:resenje2})
into the expression for canonical Hamiltonian
(\ref{eq:initialham}) we obtain effective one. It easily produces
the effective Lagrangian, which on the equations of motion for
momentum $p_\mu$, turns into
\begin{eqnarray}
\mathcal L^{eff}&=& \frac{\kappa}{2}G^{eff}_{\mu\nu}\eta^{ab}\partial_a q^\mu
\partial_b q^\nu+  \nonumber\\
 &-&\pi_\alpha
(\partial_\tau-\partial_\sigma)\left[ \eta^\alpha+(\Psi_{eff})^\alpha_\mu
q^\mu\right] \nonumber\\
&+&( \partial_\tau+\partial_\sigma)\left[ \bar\eta^{\alpha}+
(\Psi_{eff})^{\alpha}_\mu q^\mu\right] \bar \pi_{\alpha}\nonumber +\frac{1}{2\kappa} \pi_\alpha F_{eff}^{\alpha
\beta}\bar \pi_{\beta} \, .\label{eq:lag}
\end{eqnarray}
We are going to find under what conditions the initial Lagrangian
(\ref{eq:SB})  produces the effective one. We can achieve that if
we replace initial variables $x^\mu$, $\theta^\alpha$ and
$\bar\theta^\alpha$ with corresponding effective ones $q^\mu$,
$\eta^\alpha$ and $\bar\eta^\alpha$, $\Big($momenta independent
parts  of their solution (\ref{eq:resenje2})$\Big)$ and make
substitution
\begin{eqnarray}\label{eq:prelaz}
G_{\mu\nu}&\to& G_{\mu\nu}^{eff}\, ,\quad \psi^\alpha_{+\mu}\to
2(\Psi_{eff})^\alpha_{\mu}\, ,\quad F_a^{\alpha\beta}\to
F_{eff}^{\alpha\beta}\, ,\nonumber \\ B_{\mu\nu}&\to& 0\, ,
\,\,\,\,\qquad \psi^\alpha_{-\mu}\to 0\, ,\qquad\qquad\,\,
F_s^{\alpha\beta}\to 0 \, ,
\end{eqnarray}
where $G_{\mu\nu}^{eff}$, $(\Psi_{eff})^\alpha_\mu$
and $F^{\alpha\beta}_{eff}$  are  defined in (\ref{eq:effbackground}).
Note that $F_{eff}^{\alpha\beta}$ is antisymmetric by definition.

Consequently, in effective theory only $\Omega$ symmetric  parts
survive: graviton $G_{\mu\nu}$ in the NS-NS sector, gravitino
$\psi^\alpha_{+\mu}$ in NS-R sector and antisymmetric part of R-R field strength,
$F_a^{\alpha\beta}$. In our approach the effective
theory has been obtained from initial IIB one on the solution of
boundary conditions. As its $\Omega$ symmetric part, it
corresponds to the type I superstring theory, but with improved
background fields. As a consequence of boundary conditions at
$\sigma=\pi$, this is a closed string theory.

\section{\label{sec:level1}Concluding remarks}

In this paper we considered Green-Schwarz formulation of the type
IIB superstring introduced in Ref.\cite{susyNC}. Using canonical
method, following \cite{radepjc}, we derived boundary conditions
from the requirement that Hamiltonian, as time translation
generator, has well defined functional derivatives in
supercoordinates and canonically conjugated supermomenta. For
bosonic coordinates $x^\mu$ we chose Neumann boundary conditions,
while for same chirality fermionic coordinates $\theta^\alpha$ and
$\bar\theta^{\alpha}$, in accordance with
\cite{susyNC}, we chose that
$(\theta^\alpha-\bar\theta^{\alpha})|_0^\pi=0$ . As a consequence, the corresponding fermionic
canonical momenta are equal at the boundary,
$(\pi_\alpha-\bar\pi_{\alpha})|_0^\pi=0$.

All boundary conditions at string endpoints we treated as
canonical constraints. After Dirac consistency procedure,
the infinite sets of the constraints at the endpoints
can be rewritten in $\sigma$-dependent form. The constraints at
$\sigma=\pi$ and $\sigma=0$ are equal if we require $2\pi$
periodicity of all canonical variables. So, the theory obtained on
the solution of boundary conditions is closed string theory.

Solving the constraints at $\sigma=0$ we obtained the expressions
for supercoordinates  $x^\mu$, $\theta^\alpha$ and
$\bar\theta^{\alpha}$ in terms of effective ones $q^\mu$,
$\eta^\alpha$ and $\bar\eta^\alpha$ and their canonically
conjugated momenta.
The theory obtained from initial  IIB superstring theory on the
solution of constraints we will call effective theory. Effective
Lagrangian is symmetric under
orientfold projection $\Omega$ and consequently, it corresponds to
the type I closed superstring theory. It propagates in effective background obtained from the
type IIB one by substitution
(\ref{eq:prelaz}) and  (\ref{eq:effbackground}), or in terms of tensors
instead of R-R field strength by substitution (\ref{eq:sub2b1}).
The first terms in effective backgrounds are
just standard $\Omega$ projections (unoriented part of IIB
superstring theory) and second parts are our improvement.

Consequently, the type I superstring can not recognize  explicitely oriented
parts of IIB theory: NS-NS antisymmetric field $B_{\mu\nu}$,
difference between two gravitinos $\psi^\alpha_{-\mu}$ and R-R
fields: scalar $C_0$ and four rank antisymmetric tensor
$C_{\mu\nu\rho\sigma}$, but it can see  $B_{\mu\nu}$ and
$\psi^\alpha_{-\mu}$ implicitely through effective backgrounds
$G^{eff}_{\mu\nu}$, $(\Psi_{eff})^\alpha_\mu$ and
$F_{\mu\nu\rho}^{eff}$.

We can formulate our result in other way that we can express the
closed type  I superstring theory in the form of open type IIB
superstring theory, with appropriate choice of boundary conditions
(\ref{eq:gamami})-(\ref{eq:bcf1}) and appropriate relation between
background fields (\ref{eq:effbackground}). Note that type IIB background
fields are not uniquely defined by type I ones.
For fixed type I theory, there is a class of corresponding IIB
theories defined by different background $G_{\mu\nu}$,
$B_{\mu\nu}$, $\psi^\alpha_{-\mu}$ and $F_a^{\alpha\beta}$ which
produce the same $G_{\mu\nu}^{eff}$, $(\Psi_{eff})^\alpha_\mu$ and
$F_{eff}^{\alpha\beta}$.

Let us shortly discuss noncommutative properties of the open type
IIB superstring.  On the solutions of the boundary conditions
(\ref{eq:resenje2}) original string variables depend both on
effective coordinates and effective momenta. This is a source of
noncommutativity relations
\begin{eqnarray}
\{x^\mu(\sigma)\, ,x^\nu(\bar\sigma)\}&=&2
\Theta^{\mu\nu}\theta(\sigma+\bar\sigma)\, ,\nonumber\\
\{x^\mu(\sigma)\,
,\theta^\alpha(\bar\sigma)\}&=&-\Theta^{\mu\alpha}\theta(\sigma+\bar\sigma)\,
,\nonumber\\ \{\theta^\alpha(\sigma)\,
,\bar\theta^{\beta}(\bar\sigma)\}&=&\frac{1}{2}\Theta^{\alpha\beta}\theta(\sigma+\bar\sigma)\,
,
\end{eqnarray}
where $\theta(\sigma+\bar\sigma)$ is the step function. In the
case when the same chirality gravitinos $\psi^\alpha_\mu$ and
$\bar\psi^\alpha_\mu$ are equal, it follows that
$\psi^\alpha_{-\mu}=0$ and
$(\Psi_{eff})^\alpha_\mu=\psi^\alpha_\mu$, so that we reproduct
the results of Ref.\cite{susyNC}. More details about noncommutativity of
type IIB superstring we will publish elsewhere.

Therefore, the background fields odd under $\Omega$ transformation
have two roles. They are source of noncommutativity of
supercoordinates and two of them, $B_{\mu\nu}$ and
$\psi^\alpha_{-\mu}$, contribute to type I superstring background
as bilinear combinations.

\begin{acknowledgments}
Work supported in part by the Serbian Ministry of Science, under contract No. 141036.
\end{acknowledgments}

\end{document}